# DFT Perspective of Hydrogen Storage on Porous Materials

N.S. Venkataramanan, Y. Kawazoe

Institute for Materials Research (IMR)

Tohoku University,

2-1-1, Katahira, Aoba-ku, Sendai – 95035, Japan

**CONTENTS**

1.  **INTRODUCTION**

Fossil fuels represent a vital energy resource for human activities and there are growing concerns that oil reserves cannot be sustained in the face of increasing worldwide demand. The increasing level of carbon dioxide ($CO_2$) in the atmosphere produced from the burning of fossil fuels has also raised increasing concerns over the impact on global eco-systems [1]. Ongoing research efforts to address these concerns have focused on many aspects including capture and storage of $CO_2$, and the use of cleaner energy sources such as methane ($CH_4$) or methanol as alternatives to petroleum or diesel in vehicular applications. The potential use of dihydrogen ($H_2$) as an energy carrier in principle reduces or even eliminates $CO_2$ emissions entirely at the point of use. To realized "The Hydrogen Economy" [2–7] the prime goal is that of inventing safe, efficient and effective stores for $H_2$ gas, and to replace current technologies based around the compression of $H_2$ as a liquid or as a gas using cryogenic temperatures or high pressures [7]. Therefore, there is major world-wide interest in meeting the United States, Department of Energy (DOE) targets of 6.5 wt% gravimetric and 45 g $L^{-1}$ volumetric $H_2$ storage by 2010, and 9.0 wt% and 81 g $L^{-1}$ by 2015 for mobile applications [8]. For all of these concepts to be realized in the applications the key target and required technological advance is the invention of new functional materials that are highly efficient in $H_2$ reversible storage.

In the last 20 years, computer simulation studies on materials have contributed significantly to advance our understanding the mechanism and physics behind the materials, which is at the same time of fundamental scientific interest and of great technical importance [9]. This progress has been possible, on the one hand, because of improved simulation algorithms and the invention of powerful computers. Computational chemistry can be regarded as the application of chemical, mathematical, and computing skills to the solutions of chemical problems. Obtaining approximate solutions to the Schrödinger equation is the basis for most of the computational chemistry performed today. The quantum-chemical applications performed serve many times as source of inspiration for new methodological developments [10]. Among the quantum-chemical methods Density Functional Theory (DFT) is nowadays one of the most popular methods for ground state electronic structure calculation because of the favorable balance between accuracy and computational efficiency. Hence in recent years, DFT has been used to predict and realized new materials and to understand the properties of materials that can store hydrogen reversibility.

The host–guest or inclusion compound in which the lattice framework with porous (host) can accommodate the guest atoms or molecules is probably one of the most suitable hydrogen storage media [11]. This type of material belongs to the field of supramolecular chemistry, which can be defined as a chemistry beyond the molecule, referring to the organized entities of higher complexity that result from the association of two or more chemical species held together by intermolecular forces. At the present time, the role of the supramolecular organization in the design and synthesis of new materials is well recognized and assumes an increasingly important position in the design of modern materials [12]. The combination of nanomaterials as solid supports and supramolecular concepts has led to the development of hybrid materials with improved functionalities. This "heterosupramolecular" combination provides a means of bridging the gap between molecular chemistry, material science, and nanotechnology.

Among the supramolecular materials, Metal – Organic Frameworks (MOFs), Covalent Organic Frameworks (COFs) and organic porous materials find great importance due to their potential application in gas storage and catalysis [13–15]. The existence of such porous materials has been recognized for decades but it was only in the early 1990s that they were recognized as potential porous hosts for substrate inclusion *via* the formation of stable 3-D suprastructures [16–18]. 2-D and 3-D coordination polymers with metal as linkers showing permanent porosity are often named metal-organic frameworks (MOFs), but conceptually they are no different to coordination polymers which can also be regarded as crystal-engineered solids [21–23].

Physisorption of dihydrogen within a light, porous and robust material is an especially attractive option since this maximizes the possibility of highly reversible gas storage with fast kinetics and stability over multiple cycles. Although physisorption of $H_2$ in porous hosts can be highly reversible *via* changes in pressure and/or temperature, the storage involves low binding energies and isosteric heats of adsorption (typically less that 6 kJ mol$^{-1}$) and therefore cryogenic temperatures, typically 77 K, need to be used to achieve reasonable substrate uptake capacities. In contrast, chemisorption of $H_2$ involves a much higher enthalpic Dyads contribution but may also lead to slower and poorer kinetics due to the requirement for reversible cleavage and formation of the H–H bond with concomitant generation of heat [24, 25]. Hence, there is an urgent need to find materials that has energy intermediate to physisorption and chemisorption. From the thermodynamic point of view, it has been proposed, those materials with hydrogen binding energy in the range of 0.2 – 0.5 eV would be ideal for the reversible $H_2$ storage [26].

Recently computational studies, involving both electronic-structure methods (*ab initio* and density functional theory) and classical molecular mechanics (Grand Canonical Monte Carlo methods have added insight to these remarkable materials and the mechanisms of hydrogen adsorption [27]. The purpose of the review is to examine how effectively, theoretical methods have helped to elucidate the nature of interactions porous materials and the hydrogen molecules. In particular discussions were made on MOF, COF and organic porous materials, and how the state of the art DFT methods have aided to the design of new novel nanomaterials.

## 2. Metal organic frame works (MOF's) as hydrogen storage materials

### 2.1. Introduction

Metal-organic frame works are 3D structures which are comprised typically of metal center(s) onto which is bound organic donor molecules (ligands) to afford coordination polymers [28]. By judicious design and choice of metal nodes and bridging ligands, specific framework topologies can be targeted, while fictionalization and modification of the ligand and metal center's can fine-tune the electronic and chemical nature of the resultant framework surface [29]. By appropriate design of metal nodes, ligand bridges, solvent, synthetic conditions and templates, MOF materials showing porosity and open structures can be prepared (Scheme 1).

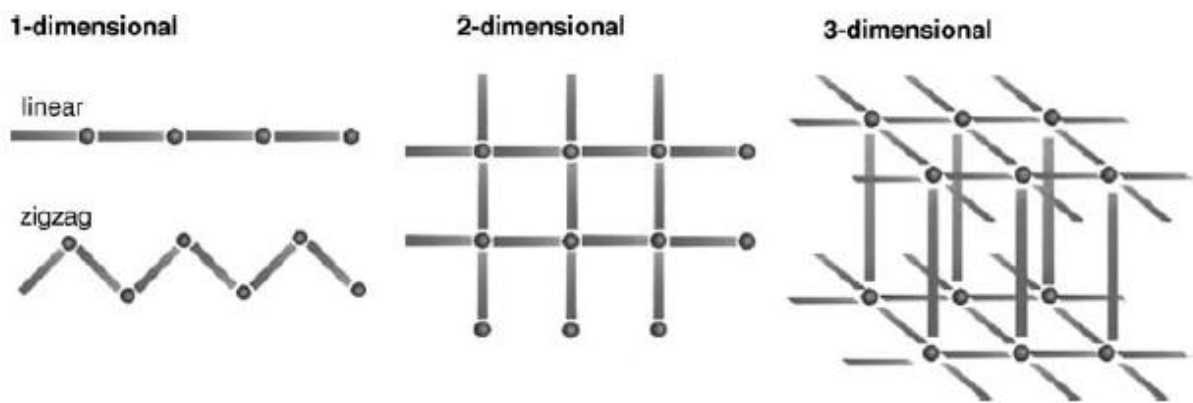

**Scheme 1.**

Compared to chemical hydrides, physisorption of hydrogen on porous materials has the advantage of fast charge–recharge processes as well as an appreciable amount of hydrogen molecules held in the pores [30]. Attention has been on the use of MOF's as hydrogen storage due to their high stability, porosity and the present of open metal sites and the initial results were encouraging. In the year 2003, Yaghi and co-workers reported the first measurements of hydrogen adsorption on an MOF: a remarkable 4.5 wt% at 77K and pressures less than 1 atm, which was latter adjusted to 1.0 wt% [31 – 34].

## 2.2 DFT as tool for designing new MOF materials

To ascertain the reason behind the attraction of hydrogen on these pore MOF's computational modeling were carried out and the results reveal that hydrogen are attracted towards the open metal sites [35]. Further studies shows that smaller pores actually take up hydrogen more effectively than very large ones [36]. Reducing the pore size allows the dihydrogen molecule to interact with multiple portions of the framework; in a smaller pore. The ideal pore size seems to be 4.5 – 5 Å , or approximately 2.8 – 3.3 Å when the van der Waal radii of the atoms composing the pore walls are excluded; this is comparable to the 2.8 Å kinetic diameter of $H_2$. This leads to optimal interaction between the dihydrogen molecule and the framework, thus maximizing the total van der Waals forces acting on dihydrogen.

Although the open metal sites are the preferential adsorption sites for hydrogen, the organic linker can play an important secondary role in increasing adsorption further [37]. Increasing the aromaticity of these organic ligands has been theoretically predicted and experimentally proven as an effective way to improve hydrogen adsorption capacity [38]. Organic linkers with aromatic fragments, such as phenylene, naphthylene, and biphenylene, are widely used in the synthesis of MOFs to form a rigid three-dimensional porous framework [39]. In the IRMOF series developed by Yaghi and coworkers, the basic structural motif of $Zn_4$(m4-O)$(CO_2)_8$ SBUs connected by aromatic phenyl containing linkers is repeated to generate a series of isostructural materials, which differ only in the central portion of the ligand as shown in **Figure 1 [31]**. Increasing the aromaticity of this central portion, from a simple phenyl ring

(MOF-5/IRMOF-1) to cyclobutylbenzene (IRMOF-6) to naphthalene (IRMOF-8) increases the hydrogen uptake dramatically, from 0.5 to 1.0 and 1.5 wt%, respectively. This result has leaded to further explore the MOF using computational methods and several suggestions to modify the organic linkers have been proposed [40 – 45].

*Insert Figure 1*

In addition to increasing the aromaticity of the organic ligands, chemical modification of the organic linkers by introducing an electron-donating group (or groups) has been suggested, based on *ab initio* calculations, as another way to further enhance framework affinity for the dihydrogen molecule. This is illustrated in the hydrogen adsorption studies of the IRMOF series. Adding one –Br, one –NH2, or four methyl groups to the central benzene ring of the linker in IRMOF-1 affords IRMOF-2, -3, and -18, respectively, while replacing the phenyl ring of bdc with a thieno-[3,2b] thiophene moiety affords IRMOF-20 [46]. The increased polarizability of the heteropolycyclic ligand improves hydrogen sorption on a molar basis in IRMOF-20 due to a stronger interaction of hydrogen with the organic linker, despite a reduction in gravimetric capacity due to the heavy sulfur atom.

*Insert Figure 2*

This idea on the modification of organic linkers has been latter extended by doping light weight alkali and alkali earth metals which helps to change the polarizability of hydrogen which improves hydrogen sorption on these materials [47 – 55]. To examine the effect of Li doping, we first investigated the adsorption energy of doping Li atom on the benzene unit of Zn-MOF-5 [56]. A model consisting of a primitive cell which represents the unit cell as shown in **Figure 2** was choosen and was fully optimized without any geometrical constraints. A summary of Li adsorption energy on Zn-MOF-5 can be seen in **Table 1**. The energy gain in attaching Li to a MOF-5 unit is called the adsorption energy (AE) and is defined as follows:

$$AE = E(\text{Li-MOF}) - E(\text{Li}) - E(\text{MOF}) \quad (1)$$

where $E(\text{Li-MOF})$, $E(\text{Li})$ and $E(\text{MOF})$ are the total energies of the unit cell containing the adsorbed Li atoms and the energy of the MOF-5 unit respectively. A positive value of $AE$ would indicate an energy gain in attaching the Li to the MOF-5 surface, and the negative value is energy lost during the process. From our calculations it is evident that doping Li atom was found to be exothermic in nature. In addition a significant change in the bond length of benzene unit was observed, which was attributed to the charge transfer from the Li cation to the MOF-5 unit. Furthermore, up on Li functionalization only a slight change in structure and bond parameters is found to occur near $OZn_4$ tetrahedra site. The calculated Li atom - benzene adsorption energy was lower than the reported values of 1.6 eV by Dixon and co-workers [57]. Such a decrease is expected as there exist effective conjugation between $OZn_4$ tetrahedra unit and linker Benzene dicarboxylic (BDC) unit. Bader charge analysis on the system revealed, that Li atom carries a +0.9e charge over it.

To understand the effect of hydrogen adsorption on the Li-functionalized MOF's we have optimized the Li-functionalized Zn-MOF-5 with hydrogen molecules near to the Li atom. Upon structural relaxation only slight change in structure and bond parameters was found to occur on the organic linker and near OZn4 tetrahedra. The hydrogen interaction or the binding energy per hydrogen molecule ($\Delta E_b$) can be defined as

$$\Delta Eb = [ET(\text{Li-MOF}) + ET(\text{H}_2) - ET(\text{Li-MOF}+\text{H}_2)] / n\text{H}_2 \quad (2)$$

where $ET(\text{Li-MOF}+\text{H}_2)$ and $ET(\text{Li-MOF})$ refer to the total energy of the Li- functionalized Zn-MOF with and without hydrogen molecule respectively, while the $ET(\text{H}_2)$ refers to the total energy of the free hydrogen molecule and "n" is the number of hydrogen molecules. In the current work we have added one to four hydrogen molecules near the Li center. The calculated structural parameters and the binding energies are provided in **Table 2**.

For the first $H_2$, the interaction energy is 0.213 eV, with an intermolecular distance of 2.153 Å. This binding energy per atom was close to the value reported for systems that can be used for an ideal reversible hydrogen storage system [58]. The orientation of hydrogen is in T-shape, with a H-H bond distance of 0.760 Å, which corresponds to a very small change compared to the 0.750 Å bond length in a free hydrogen molecule. To ascertain the reason of such interaction we plotted the Electrostatic potential and the charge density difference of model system as shown in **Figure 3** with Li-functionalized benzene molecule. It is evident from the figure a charge decrease could be adsorbed on the hydrogen molecule upon the adsorption of hydrogen. This indicates that the Li cation holds the $H_2$ molecules by a charge – quadruple and charge – induced dipole interaction. When the second hydrogen is introduced, the Li – $H_2$ distance increases to 2.124 Å, while the binding energy per $H_2$ molecule gets reduced to 0.209 eV. To know the number of hydrogen molecules a Li cation can hold, we doped the third and fourth $H_2$ near the Li cation. With the introduction of third $H_2$ the Li – $H_2$ distance increases along with a decreases in the interaction energy value. A noticeable feature is the H – H distance which decreases with the increase in the number of $H_2$. When the fourth hydrogen is introduced near the Li cation, three $H_2$ are place near the Li atom and the other $H_2$ molecule is moved away to a nonbonding distance of 4.036 Å. Thus each Li cation can hold up to 3 hydrogen molecules in a quasi molecular form. Furthermore, each BDC unit can adsorb one Li atom on each face and can maximize its storage to 6 hydrogen molecules per BDC unit. Further the existence of several metal sites may increase the storage capacity of the Li-functionalized MOF's.

*Insert Figure 3*

To know the possibility of extending the Li doping to other IRMOF-5, we studied the adsorption of Li cation on IRMOF-5. We have selected metals ( M= Fe, Co, Ni, Cu and Zn) from the same column of the periodic table by replacing Zn centers. Structural relaxation without Li atom shows only a slight change in volume to occur up on the change in metal sites. However, up on doping with Li atom a considerable change in shape and structure was observed for other materials except Zn. **Table 3** contains structural information for the linker and the benzene-Li mean distance. One can observe that the linker unit benzene remains practically unchanged in all the studied compounds.

In **Table 4.** we have collected the M-O1, C1-O1, C1-C2 bond lengths, O1-M-O2 bond angles and the adsorption energy of Li for the optimized structure of M-MOF-5 system. From the table it is evident that a significant structural change is observed around the OM4 tetrahedra unit and linker BDC unit. The metal-metal distance deviation is shorter in the case of Zn, and very high reduction in distance is observed for iron system. Further, calculated adsorption energy for these compounds don't show any regular trend with the metals. These results suggest that adsorption energy of Li not only depend on the metal centers but are greatly influenced by the structural change and the volume of the system that changes upon Li doping.

### 2.3. Out look

The use of DFT theory in this flourishing area of interdisciplinary research has helped researchers to understand the mechanism of hydrogen adsorption. Recent discoveries have led to a growing understanding of the underlying structure–property relationships for this class of material although much remains to be uncovered. Fixing $Li^+$ on the internal surface of MOFs can potentially stabilize a partially exposed $Li^+$, and some recent experimental reports have indicated that this is a viable route to improving $H_2$ adsorption enthalpy and also the gravimetric density on these materials. However, the challenge still remains to design new functional and robust materials incorporating higher pore volumes and specific multi-functional groups.

### 3. Covalent Organic Frameworks (COF's) as hydrogen storage materials

### 3.1. Introduction

Covalent Organic Frameworks (COF) are similar to those of MOFs, in terms of structural integrity, however, lags the presence of metal sites which are responsible for the high molecular weight of MOFs [59]. This offers addition advantage of storing high gravimetric weigh percentage of gases in the COFs'. Moreover, functional group modification and subsequent purification of these organic molecules is readily achievable in solution than the heterogeneous chemical transformation in the MOFs. In addition, COFs have high thermostabilities of the up to $500^oC$ that makes these compounds potential candidates for use in industrial processes. In these materials the organic building units are held together by strong covalent bonds (C-C, C-O, B-O, and Si-C) rather than metal ions [60]. Most of the above-mentioned low-density materials present a relatively high surface area and high microporosity. Henceforth, in recent years attention has been focused to use them as hydrogen storage media.

Recently, Yaghi et al. developed the synthetic strategy for COF materials [61]. The building blocks for COFs were 1,4-benzenediboronic acid (BDBA) and hexahydroxytriphenylene (HHTP). Under carefully chosen reaction conditions, BDBA reacts with itself, and the boronic acid moieties condense to planar boroxine rings ($(RB)_3O_3$, (**Scheme 2**) [62]. For the synthesis, the selection of the right solvent seems to be important. To foster the formation of a uniform and highly ordered structure, solvents are chosen wherein the reactants are poorly soluble. This approach slows down the reversible condensation. Furthermore, the reactions are carried out in sealed pyrex tubes, again to slow down the reversible process and minimize defects by self-healing. COF-1 was isolated as a microcrystalline substance in high yield. Powder X-ray diffraction patterns reveal a high structural order, whereas single interlayers are stacked in a staggered arrangement. Solvent molecules are enclosed inside the 15-Å-diameter pores and can be removed at $200^0C$ without collapse of the crystalline structure. COF-1 has a surface area of 711 $m^2\ g^{-1}$ and a pore volume of 0.32 $cm^3$ g-1. These values are comparable to those of porous zeolite and carbon-based materials.

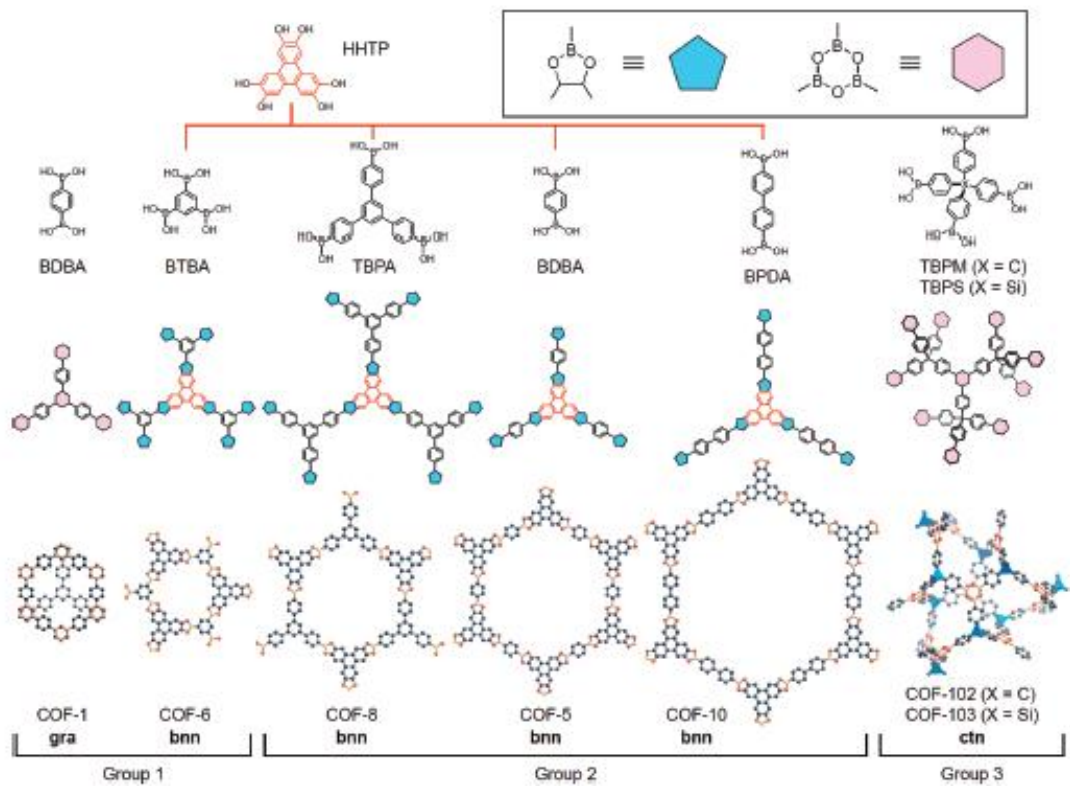

**Scheme 2.** Schematic representation of synthetic procedures for the 2D-and 3D – COF materials

Recently, Yaghi and co-workers expanded the networking scope to three-dimensional space [63 – 65]. Using tetrahedral tetraboronic acids as precursors, 3D frameworks (COF-102, COF-103, COF-105, and COF-108) resulted from condensation or co-condensation with HHTP. Although the yields are lower than for the 2D frameworks, the 3D COFs display excellent physicochemical properties. Based on these synthetic strategies different diboronic acids were condensed with tetraols and new class of COFs was achieved.

## 3.2. Structural design for hydrogen storage on COF's using DFT

Most porous COFs are extremely lightweight, many with higher surface area and lower crystal density [66]. In most cases, the low density serves to reduce the volumetric hydrogen uptake of the COF material despite its high-gravimetric uptake. Nevertheless, in terms of absolute adsorption, there exist some COF materials with both gravimetric and volumetric hydrogen uptake at 77K which surpass the 2010 DOE targets for hydrogen storage [67 – 70]. It would seem very promising to store hydrogen in porous COFs at 77K and high pressure due to the week physisorption of hydrogen on these materials [71]. However, the cost and weight of the cryogenic pressure vessel precludes a practical on-board application, and porous COFs with high-hydrogen uptake near ambient temperature are badly needed.

On this account, modeling studies were carried out to increase the adsorption enthalpy of hydrogen. Typically, doping is the concept used to alter the physicochemical properties of many materials [72]. A considerable number of computational studies regarding hydrogen molecules interacting with isorecticular COFs have recently been published [73 – 76]. The interaction energies and the corresponding geometries have been calculated at diverse levels of theory. In the past, the interaction energy of $H_2$ with the organic linkers has been determined to be 0.03–0.05 eV. This energy is very similar to the theoretically calculated energy of interaction of a hydrogen molecule with benzene.

To investigate the adsorption sites and the adsorption energies of hydrogen molecules, Kang *et al* carried out DFT studies on the three dimensional COFs ( COF-102, COF-103, COF-105, and COF-108). The best adsorption site was found on the oxygen atom of the $C_2O_2B$ ring with a binding energy of 0.06 eV [77]. They employed Li, Mg, $Li^+$, $Mg^{2+}$ and Ti as dopant and found Li and Mg prefers to exist as isolated elements on the COFs. Further, they increase the binding energy of these materials. Each decorated $Li^+$ and $Mg^{2+}$ can adsorb up to three and six $H_2$ molecules with the average binding energy of 0.28 and 0.30 eV, respectively. Contrary to their results Assfour and Seifert, based on dispersion-corrected density functional based tight binding theory (DC-DFTB) and molecular dynamics proposed that benzene and the aromatic rings present in the COF materials were the site of adsorption of dihydrogen molecule [78].

In 2008 Choi et al proposed a new design called pillared COFs, which consists of pyridine as linker [79]. The pyridine nitrogen atom was located close to the boron atom either to form a covalent bond or to undergo a physisorption interaction helping to form a pillared COF material. Zang et al proposed Ca as the dopant and by using density functional theory and second-order Møller–Plesset perturbation calculations show that the $H_2$ binding in COF systems can be significantly increased to 5.7 wt % of gravimetric density [80]. Ganz et al carried out studies for the model systems of COF-1 and IRMOF-1 and suggested that experimental studies on IRMOF has been optimized, however, COF has to be studies further [81]. They suggest that it is possible to store significant amount of hydrogen by spillover process and the significant strain will result from shrinkage of the linker molecules as H atoms are loaded onto the crystals.

A two-step doping strategy for chemical modification of COFs was proposed by Ihm et al.[82]. In which the first step involves the replacement of carbon on the organic units by the boron atom. The B substitution for C in the benzene ring of COFs is an endothermic reaction, which can be done under highly nonequilibrium conditions. The second step involves the doping of metal atoms such as Sc, Ti and Ca which acts as trapping centers for dihydrogen molecules on the B-doped COFs. The clustering of Sc, Ti and Ca were suppressed on B-doped COFs and each Sc and Ti atom bound four dihydrogen molecules by a Kubas interaction. On the other hand, Ca adsorbs six dihydrogen molecules by the polarization of $H_2$ molecules under the electric field that exists on Ca atom. They propose a hydrogen uptake of about 6.5 wt % on the COF-1 modified framework.

To explore the effect of Li and Ca metal dopent on the COFs, Jena et al studied their adsorption on the COF-10, which can be synthesized by the cocondensation reaction between 2,3,6,7,10,11-hexahydroxy triphenylene (HHTP) and 4,4-biphenyldiboronic acid (BPDA) [70]. The cluster unit used for their study is shown in **Figure 4.** The best site of dopant atoms were on top of the benzene ring. The binding energy of Li atom on COF-10 substrate is found to be about 1.0 eV and each Li atom can adsorb up to three H2 molecules. However, at high concentration, Li atoms cluster and, consequently, their hydrogen storage capacity were reduced due to steric hindrance between H2 molecules. On the other hand, due to charge transfer from Li to the substrate, O sites provide additional enhancement for hydrogen adsorption. With increasing concentration of doped metal atoms, the COF-10 substrate provides an additional platform for storing hydrogen. To avoid the clustering of the dopants, Sun et al proposed a new type of COF by incorporating a tetrazolide group into porous materials. The proposed structure provides 14 binding sites for hydrogen molecules with modest interaction energies with the predicted gravimetric density of 4.5 wt %.

*Insert Figure 4*

### 3.3. Multiscale theoretical Studies

Due to the failure of DFT methods to predicting the accurate binding energies in the weak binding systems, classical methods were adopted to understand the properties of such materials. Among them the Grand-canonical ensemble Monte Carlo (GCMC) is a very versatile and powerful technique that explicitly accounts for density fluctuations at fixed volume and temperature [83]. In addition to predicting macroscopic phenomena such as adsorption isotherms and heats of adsorption, GCMC simulations can also provide a detailed picture on the molecular scale, as the positions and potential energies of all adsorbate molecules are known over the duration of the simulation. This information can be further analyzed to get information about energetic interactions, preferential siting, and ultimately the adsorption mechanism. Potential energy maps can be easily determined by placing an adsorbate molecule at positions throughout the MOF framework and calculating the adsorbate/framework interaction, i.e. without conducting a GCMC simulation [84]. These potential maps illustrate the regions in the framework with the lowest interaction energy and therefore the preferential adsorption sites. Whereas the potential energy maps show the distribution of potential energies in the framework, they do not contain information on how these sites are filled up with increasing loading. This information is obtained by determining energy histograms during a GCMC simulation. Especially if determined for several pressures, these distributions contain valuable information on how quickly specific sites saturate and the relative importance of individual interaction sites. On this aspect, GCMC simulations were carried out extensively on the various systems including that on hydrogen storage.

In the multiscale theoretical method, several research groups have studies the gas storage properties of diverse COF materials of different structural and pore size [85 – 89]. The first report on the dihydrogen uptake by COFs from first-principles based GCMC simulations was reported by Goddard et al [90]. Their predicted uptake was comparable with the experimental results. COF-105 were capable of storing about 10.0 wt% of dihydrogen reversible at 77 K while and COF-108 was capable of 18.9 wt %. Latter Froudakis and coworkers showed COF-18 is capable of storing 21 wt% of dihydrogen by carrying a GCMC simulation, for which the force fields are derived from higher accuracy MP2 perturbation theory [91]. To improve the storage characteristics, they introduced lithium near the alkoxide groups in the COF-105. The hydrogen binding energy was found to improve and the gravimetric uptake for the system reaches 6 wt % in the near room temperature. Furthermore, they designed new COF materials by introducing phenyl groups and keeping the cnt network. GCMC simulations on these materials reveal that they are capable of storing 25 wt % and 6 wt% of dihydrogen at 77 K and at room temperature [92, 93]. The uptake capacity of nondoped and Li-doped COF-202 were later studied by Wang et al [94]. Their predictions show that that the total gravimetric and volumetric uptakes of hydrogen in the Li-doped COF-202 reach 4.39 wt % and 25.86 g/L at $T = 298$ K and $p = 100$ bar, respectively, where the weight percent of Li equals to 7.90 wt %. Cao and coworkers proposed 4 new COF materials which reflects Diamond like structure whose structure are varied by changing the chain length using the benzene unit [95]. The highest uptake was found in PAF - 304, with gravimetric density reaching 6.53 wt%.

### 3.4. Outlook

The development of accurate DFT methods and their use in deriving force field parameters has provided a new approach in designing new materials. These simulated results may provide a guide for the experimentalist to design novel materials and to understand their macroscopic properties at the molecular level. This combined scheme for the development of novel porous materials is expected to play an increasingly important role in the future, and an increasing number of novel materials for gas storage are likely to be developed using this approach.

## 4.    Organic Nano-Porous Materials

### 4.1.    Introduction

Porous networks polymers such as COFs and MOFs can show a range of functionalities and pore size, agonizing synthetic methods make them difficult to study. There exist molecules which are discrete but are porous in the solid state. Such molecules, may either crystalline or amorphous, and can be categorized as either intrinsically porous or extrinsically porous [96]. There porous are capable of storing gases reversibly. Hence in recent years attention has been focused on these systems and hydrogen has been successfully store in such systems. In addition the use of organic systems provides four fold remunerations. Firstly, the discrete molecules are soluble in common organic solvents or in water, which is advantageous in terms of their synthesis to desired topology. Secondly, due to their solubility, two different functionality can be incorporated in to a single crystalline structure. Thirdly, post synthetic modification can be readily is readily achieved in solutions compared to the heterogeneous system. Fourthly, molecular crystals — and amorphous molecular solids — are not interconnected by strong covalent or coordination bonds [97]. This could in turn allow the cooperative trapping of specific guests. Due to this addition advantageous compared to MOFs and COFs organic porous solids, many research groups started to focus on the gas sorption ability of these systems.

Recently Atwood *et al.* demonstrated with the pure organic compound, *p*-tert-butylcalix[4]arene (TBC), readily and reversibility absorb gases such as $H_2$, $N_2$, $O_2$, $CO_2$ and $CH_4$. The preliminary results indicate that the TBC has the capability to store hydrogen up to 0.6 wt% [98 – 100]. The packing mode obtained for the crystals shows existence of each cup-shaped host molecules facing another host molecule in the adjacent layer, to form a relatively large cavity of 270 Å$^3$. In addition the cavity posses a π–rich character defined by the four aromatic rings, which is sufficient enough to store guest molecules. Latter similar cages structures were constructed and were tested for their gas sorption properties.

**4.2.    Organic materials as hydrogen storage materials**

In the computational part, molecular dynamics study on the interaction of $CO_2/H_2$ with TBC was studied using potential of mean force and free energy of perturbation approach by Dang *et al.* The found that $CO_2/H_2$ have favorable interaction with TBC, with the $CO_2$ interaction being considerably greater [101]. Recently Alavi and Ripmesster studied the gas adsorption properties of the calix[4]arene by performing a molecular dynamics study [102]. The inclusion energy calculated favors for xenon to be the better guest molecules and they occupy the cage and interstitial sites. $H_2$ binding due to charge-induced dipole interactions with light metal atoms has been an important topic of chemistry for many decades.

In our lab, we have carried out combinatorial studies on the dihydrogen adsorption on the TBC and the Li-functionalized TBC molecule [103, 104]. The structure of TBC was fully optimized without any geometrical constraints. The optimized structure is shown in F**igure 5 (a)**. The lowest energy structure has a geometry in which phenoxy hydrogen was bonded in one direction and the t-butyl groups are oriented in the opposite direction.  Li absorption on the TBC can take place at four different sites:  by replacing the hydrogen on phenol to form an alkoxy salt (O –Li), as cation at the center of the four phenoxy group and on the walls of the benzene ring making a Li-benzene π-complex.

*Insert Figure 5*

Figures **5 (b-e)** shows the four different sites considered for the Li absorption. The preferred position of Li is found to be on the inside wall of benzene(Fig. **5 (d)** ), with a binding energy of 0.714 eV, which was calculated from the energy difference between total energy of Li-functionalized calixarene and TBC. The energy of the neutral compound (**Fig. 5 (b)**) was about 0.04 eV higher in energy, while the least stable structure was the one in which the Li atom is bonded on the outside wall of the benzene ring (**Fig. 5 (e)**). In our further studies, we consider the structure with Li atom bounded on inside wall complex (LTBC) alone, as the Li atom doped will be rigid, and the anions can occupy the pore spaces. In addition, this would avoid Li clusters in these systems. Li doping on TBC molecule induces, increase in bond lengths of the benzene unit, indicating a charge transfer occurs from the Li atom to the TBC molecule.

We first studied the hydrogen molecule uptake for TBC molecules. **Figure 6** shows the optimized structure for the TBC molecule with one (**Fig 6 (a,b)**) and two molecular hydrogen (**Fig 6(c)**). The first hydrogen molecule stays at a distance of 4.75 Å from the bottom and center of the four phenoxy units and is oriented parallel to the phenoxy group. Another important feature is the minimum energy structure has one t-butyl group orient in the direction of the phenoxy hydrogen. Upon doping additional hydrogen, the first hydrogen molecule is pulled inside and resides at a distance of 4.58 Å, while the second hydrogen molecule is partially placed inside the calixarene cavity at a distance of 6.67 Å. The hydrogen molecules in the optimized structures have bond distance of 0.750 Å, which are the same obtained for isolated molecular hydrogen optimized with PW91 GGA method. This reflects the absence of any interaction between the TBC molecule and hydrogen molecule.

*Insert Figure 6*

The calculated binding energies and bond lengths for the hydrogen molecules are provided in **Table 5**. The preferred position for the first $H_2$ molecule is found to be at 2.085 Å away from the Li atom at a distance of 4.16 Å from the bottom and center of phenoxy unit. The $H_2$ binding energy per hydrogen molecule was calculated using the expression

$$BE/H_2 = (E(LTBC+nH_2) - E(LTBC) - E(nH_2))/n,$$

where $E(LTBC+H_2)$ was the total energy of Li-doped TBC containing 'n' number of hydrogen and $E(LTBC)$ total energy of the Li-doped TBC.

In the ground-state configuration, hydrogen atoms are bound in nearly molecular form with H – H bond length of 0.758 Å, with hydrogen molecule occupying the parallel position to the Li atom. Upon introducing the second and third hydrogen molecules the binding energy per molecule reduced to 0.292 eV and 0.248 eV for the second and third hydrogen molecule. Another noticeable feature is the increase in the distance between the Li atom and the hydrogen with the addition of successive hydrogen. Upon reacting fourth hydrogen, one hydrogen molecule was found to fly away from the Li atom and found to be at a distance of 2.97 Å and was inside the cavity of the LTBC. Thus the maximum number of hydrogen molecules bound by Li atom is three, while the LTBC can hold four inside its cavity. Further, each TBC molecule can be functionalized with 5 Li atom on it's out and inside rim.

*Insert Figure 7*

To investigate the stability of Li –functionalized calixarene, the pair distribution function (PDF) which is the mean distance between the benzene ring (on which Li atom is functionalized) and the Li atom, over a temperature rage of 20 to 300 K, were calculated using *ab initio* molecular dynamics. To attain the equilibration, the system was allowed to reach 1500 steps of

2 fs, after which the coordinates are analyzed to investigate its stability. As seen in **Figure 7(a),** the Li atom stays to the benzene ring for the entire temperature range. Recently Ahuja *et al.* have used the same methodology in Li decorated MOF's. Following to the study, LTBC stability with hydrogen molecules were also measured by calculating mean distance between the Li atom and the center of the hydrogen molecules for the system with four $H_2$ molecules inside its cavity [105]. The results obtained are shown in **Figure 7(b, c)**. It is clean from the Figure that the Li atom was intact to the benzene ring until 200 K. upon further increase in temperature results in the decomposition of the complex. Hence for the Li-$H_2$ system, we have calculated the stability until 200 K.

Recently, we have studied the adamantane (ADM) systems for hydrogen storage due to their ability to form porous structure and their easy capability to Lithiate them [106]. In all our calculations, it was observed that the binding energy of hydrogen molecules on ADM was very small, *i.e.,* on the order of ~0.001 eV, which is out of the energy range, from -0.1 to -0.2 eV, desirable for reversible $H_2$ adsorption/desorption near room temperature for hydrogen storage applications. Hence, pristine diamondoids are not considered to be good candidates for hydrogen storage applications. However, experimentally it has been shown that by treating the surfaces of diamondoids with chemical solutions, the hydrogen atoms of diamondoids can be selectively replaced by other compounds or elements, resulting in significant changes in their electronic structures. Among the experimentally formed diamondoid-based complexes, one modified with Li/Li+ can attract attention for hydrogen storage applications and it would be worthwhile to consider their storage properties in detail [107]. The wide range of experimental results for hydrogen storage improvements after Li doping, from less than 1 wt% to several tens of weight percentages at moderate pressures and temperatures, are mainly attributed to sample preparation issues. Since ADM has four acidic hydrogen atoms, it can possibly be modified with four Li atoms/ions. Our calculations show that in adamantane molecules, more than one acidic hydrogen cannot be replaced by a Li ion (Li+) because the structure of adamantane is significantly deformed such that the binding energy of the second substituted Li ion becomes positive, +0.15 eV .

**Table 6** shows the binding energy of Li/Li+ in ADM.Li$_m$ (m=1-4) and ADM.Li+. The trend of the changes of binding energies of Li to ADM.Li$_m$ are observed to be the same using different methods, PW91, PBE, M05-2X, and MP2; by increasing the number of Li atoms, the binding energy of Li to ADM.Li$_m$ decreases. Furthermore, our calculations show that the binding energy of Li in ADM.Li is almost same as the binding energy of Li+ in the ADM.Li+ complex. To confirm the stability of ADM.Li$_m$/Li+, we have calculated the vibrational spectra of our designed structures. The obtained results show the absence of any imaginary frequency, indicating that the optimized structures are real minima. We also performed MD simulations at 400 K for 48 ps and observed no detachment of Li/Li+ from the structures and no Li aggregation on an individual ADM.Lim complex. The possibility of formation of ADM.Na and ADM.K complexes was also considered, and the binding energy of Li to ADM was found to be larger than that of Na (-0.86 eV) or K (-0.76 eV) atoms to ADM. Hence, we focused our examination on just ADM.Li$_m$/Li+ complexes.

*Insert Figure 8*

To consider the nature of Li/Li$^+$ bonding in ADM.Li/Li$^+$, we performed excess and depletion charge analysis and plotted their total and projected density of states, as shown in **Fig. 8**. The projected density of states was plotted for the Li/Li$^+$ and the host carbon atoms. From excess and depletion charge analysis, in ADM.Li, Li was found to become positively charged by donating its 2s electron to the cluster, particularly the host carbon atom and all the hydrogen atoms in the complex. In ADM.Li+, Li+ becomes less positively charged by accepting electrons, mainly from the hydrogen atoms of the cluster. From the projected density of state calculations, the electrons donated to Li+ were observed to move not only to its 2s orbital, but also to its 2py orbital. In the ADM.Li complex, the 2s orbital of Li is partially hybridized with the 2pz orbital of the host carbon atom (see PDOS peaks at -0.25 eV) and makes a sp3-like bond with it. In ADM.Li+, the 2s and 2py orbitals of Li+ are partially hybridized with the 2pz orbital of the host carbon atom. From our calculations, it is concluded that the bonding nature of Li/Li+ in ADM.Li/Li+ is predominantly ionic and partially covalent.

Our energy calculations show that the dimer configuration is much more stable than the chain structure. Previously Xue and Mansoori have found that ADM.Na+ complexes are self-assembled like a molecular crystal by performing MD simulations on 125 ADM.Na+ complexes [108]. The vacant spaces between the complexes make it possible to store hydrogen molecules in high gravimetric weight percentages. Moreover, recently, experimentalists have been able to prepare the positively charged alkali metal doped MOF systems via electrochemical reduction. They proved that the hydrogen storage capacity increases after introducing the alkali metal charge cations. Our studied system is an organic molecule, which can be easily charged electrochemically than MOFs. Hence we believe that ADM.Li+ is a superior to ADM.Li as a candidate for hydrogen storage.

The next step of our study was to consider the hydrogen storage property of ADM.Li+ and ADM.Li$_m$ complexes. As summarized in **Table 7**, each Li/Li+ site adsorbs a maximum of five $H_2$ molecules. Therefore, it can be predicted that the gravimetric weight percentage of hydrogen storage for ADM.Li$^+$ is ~7.0 % and between 7.0-20.0 % for ADM.Li$_m$ complexes, if experimentalists can find a way to prevent their clustering. If clustering occurs the storage properties of ADM.Li$_m$ will be less than the above mentioned value. From the table, it is seen that in our case study, both pure and hybrid functional provide binding energies close to the more accurate MP2 method. It is observed that the binding energies of hydrogen molecules on ADM.Li$_m$/Li$^+$ are on the order of -0.1 to -0.23 eV, which is very good for hydrogen storage applications. Furthermore, calculated binding energies for the cationic Li$^+$ are higher than those of the neutral Li-doped ADM system. The calculated Li-H distance increases with an increase in the number of hydrogen molecules. The small changes in binding energy of hydrogen molecules can be attributed to various reasons: the amount of positive charges on Li, the distances between H$_2$ molecules and Li, the Li−C bond distance, the interaction between the hydrogen molecules, the strength of hybridization of hydrogen molecules with Li, etc.

It is observed that the adsorbed H$_2$ molecules are located at distances of ~2.1 Å of Li/Li$^+$. It is also seen that the first H$_2$ is adsorbed on top of the Li/Li$^+$, but when the second, third, or fourth H$_2$ molecules are adsorbed, they prefer to move to the lateral side of Li/Li$^+$. When the hydrogen molecules are adsorbed on the ADM.Li/Li$^+$ complexes, the position of Li/Li$^+$ changes between their two local minima with a straight or tilted Li/Li$^+$−C bond in order to reduce the steric repulsion between the adsorbed hydrogen molecules and the hydrogen atoms of the ADM.Li/Li$^+$ structures.

***Insert Figure 9***

The excess and depletion charge analysis, positively charged Li was shown to polarize H$_2$ molecules under its induced electric field. Due to this polarization, there is a small bond elongation in $H_2$ bond distances, as seen from **Table 7**, but no dissociation of $H_2$ molecules. Density of states analysis indicated that when the number of $H_2$ molecules increases from one to four, they start to interact with each other such that the states related to $H_2$ molecules are broadened, between ~-8.0 eV and -10.0 eV. To better understand how the electric filed induced by positively charged Li affects the binding energies of $H_2$ molecules, we plotted the magnitude of the induced electric field along the $Li/Li^+-C$ bond for $ADM.Li/Li^+$, shown in **Fig 9**. As shown by the figure, the amount of generated electric field at the center of adsorbed H2 molecule on $Li/Li^+$ is on the order of 2.1/3.4 V/Å. The polarizability of a hydrogen molecule along ($\alpha k$) and perpendicular ($\alpha \perp$) to the hydrogen molecule axis in an external electric field are 6.3 a.u. and 4.85 a.u., respectively. The adsorption energy of hydrogen in such electric fields is estimated to close to -0.11to -0.29 eV. These values are very close to ones reported in **Table 7**. Therefore, it is expected that the electrostatic interactions between $H_2$ molecules and Li make larger contributions to the binding energy of $H_2$ molecules on ADM.Li/Li+ complexes than their hybridizations.

*Insert Figure 10*

To investigate the thermodynamics of adsorption of $H_2$ molecules on ADM.Li and ADM.Li+, the occupation number of $H_2$ molecules per site (Li/Li+ atom) was calculated as a function of the pressure and temperature. **Figure 10** shows the occupation number of $H_2$ molecules on the Li and Li+ atoms as a function of the pressure and temperature where the experimental chemical potential of $H_2$ gas and the calculated binding energy obtained from MP2 calculations were used. The occupation number of $H_2$ molecules at 150 K and 30 atm in both cases is ~4. This is attributed to the Gibbs factor for the binding of four $H_2$ molecule, which dominates at 150 K and 30 atm (μ = −0.08 eV, ϱ4 on the Li and Li+ is -0.12 and -0.17 eV, respectively). The number goes to zero at room temperature (μ =∼ −0.32 eV). Therefore, this analysis shows that the ADM.Li/Li+ structures we suggest have considerable potential as high-capacity hydrogen storage materials.

## 5. Final Comments

In this chapter, the physisorption of hydrogen molecules in porous materials as possible hydrogen storage systems has been reviewed. Owing to the weak interaction between $H_2$ molecules and the adsorbent, high storage capacities are typically reached only at cryogenic temperature. Different classes of porous materials possessing different structure and composition have been designed for hydrogen storage applications using computational methods and especially with the aid of DFT methods. The adsorption energies for hydrogen in different porous materials have been increases by the doping of light weight alkali and alkali earth metals. *Ab initio* molecular dynamics has been carried out to know the stability of the newly functionalized materials. GCMC methods have been employed to know the gravimetric and volumetric uptake percentage of the newly functionalized materials. Therefore, the combined approach provides a better understanding and designing new materials to operate at near room temperature for the reversible hydrogen storage application.

**Acknowledgement**

We thank the New Energy and Industrial Technology Development Organization (NEDO) for funding the project "Advanced Fundamental Research Project on Hydrogen Storage Materials". The authors thank the crew of the Center for Computational Materials Science at

Institute for Materials Research, Tohoku University, for their continuous support of the HITACHI SR11000 supercomputing facility. NSV thanks his wife Suvitha for her help and support during the preparation of this book chapter.

**Reference**

1. K.S. Deffeyes, Beyond Oil: The View from Hubbert's Peak Hill and Wang, New York (2005).

2. M. Hirscher, Handbook of Hydrogen Storage: New Materials for Future Energy Storage, Wiley-VCH Verlang GmbH & Co. KGaA, Winheim (2010).

3. L. Schlapbach and A. Züttel, *Nature* 414, pp. 353-358 (2001).

4. L. Schlapbach, *MRS Bull. 27*, 675. (2002) and other articles in this special issue.

5. R. H. Jones, G. J. Thomas, Materials for the Hydrogen Economy, CRC press, Boca Raton (2008).

6. R. B. Gupta, Hydrogen Fuel : Production, Transport, and Storage, CRC press, Boca Raton (2009).

7. A. Züttel, A. Borgschulte, L. Schlapbach, Hydrogen as a Future Energy Carrier, Wiley-VCH Verlang GmbH & Co. KGaA, Winheim (2008).

8. http://www1.eere.energy.gov/hydrogenandfuelcells/pdfs/program_plan2010.pdf

9. E. G. Lewars, Modeling Marvels: Computational Anticipation of Novel Molecules, Springer Science (2008).

10. G. Petrone, G. Cammarata, Recent Advances in Modelling and Simulation, I-Tech Education (2008).

11. Jerry L. Atwood and Jonathan W. Steed, Organic nanostructures, Wiley-VCH Verlang GmbH & Co. KGaA, Winheim (2008).

12. J.-M. Lehn, Supramolecular Chemistry Wiley-VCH Verlang GmbH & Co. KGaA, Winheim. (1995).

13. T. Duren, Y.-S. Baeb, R. Q. Snurrb, *Chem. Soc. Rev*. 38, 1237 (2009).

14. M. Nagaoka, Y. Ohta, H. Hitomi, *Coord. Chem. Rev*. 251, 2522 (2007).

15. L.R. Macgillivray, Metal-organic frameworks : Design and Application, John Wiley & Sons, New Jeresy (2010).

16. Jonathan W. Steed, Jonathan W. Steed, Karl J. Wallace, Core Concepts in Supramolecular Chemistry and Nanochemistry, Wiley-VCH Verlag GmbH, Boschstr (2007).

17. H-J. Schneider, *Angew. Chem. Int. Ed*. 48, 3924 (2009).

18. G. Wenz, *Angew. Chem. Int. Ed*. 33, 803 (1994).

19. J.-P. Sauvage, Perspectives in Supramolecular Chemistry: Transition Metals in Supramolecular Chemistry, John Wiley & Sons, Chichester (1999).

20. R.W. Corkery, *Curr. Opin. Colloid In*. 13, 288 (2008).

21. A. Erxleben, *Coord. Chem. Rev*. 246, 203 (2003).

22. A.M. Beatty, *Coord. Chem. Rev*. 246, 131 (2003).

23. J.L.C. Rowsell, O.M. Yaghi, *Micropor. Mesopor. Mat*. 73, 3 (2004).

24. N.S. Venkataramanan, R. Sahara, H. Mizuseki, Y. Kawazoe, *J. Phys. Chem. A*. 114, 5049 (2010).

25. J. Roques, C. Lacaze-Dufaure, C. Mijoule, *J. Chem. Theory Comput*, 3, 878 (2007).

26. N.S. Venkataramanan, H. Mizuseki, Y. Kawazoe, *Nano*, 4, 253 (2009).

27. S. S. Han, J. L. Mendoza-Cortés, W. A. Goddard III, *Chem. Soc. Rev*, **38**, 1460 (2009).

28. A. M. Spokoyny, D. Kim, A. Sumrein, C. A. Mirkin, *Chem. Soc. Rev*, **38**, 1218 (2009).

29. M. O'Keeffe, *Chem. Soc. Rev*, **38**, 1215 (2009).

30. L J. Murray, M. Dincă, J. R. Long, *Chem. Soc. Rev*, **38**, 1294 (2009).

31. N. L. Rosi, M. Eddaoudi, D. T. Vodak, J. Eckert, M. O'Keeffe, O. M. Yaghi, *Science*, *300*, 1127 (2003).

32. J. L. C. Rowsell, A. R. Millward, K. S. Park, O. M. Yaghi, *J. Am. Chem. Soc.*, *126*, 5666 (2004).

33. B. Chen, D. S. Contreras, N. W. Ockwig, O. M. Yaghi, *Angew. Chem. Int. Ed*, *44*, 4745 (2005).

34. J. L. C. Rowsell, O. M. Yaghi, *Angew. Chem. Int. Ed*, *44*, 4670 (2005).

35. F.M. Mulder, T.J. Dingermans, M. Wagemaker, G.J. Kearely, *Chem. Phys.* 317, 113 (2005).

36. T. Düren, Y.-S. Bae, R. Q. Snurr, *Chem. Soc. Rev*, **38**, 1237 (2009).

37. A. Kuc, T. Heine, G. Seifert, H.A. Duarte, *Theo. Chem. Acc*, 120, 543 (2008).

38. J. L. C. Rowsell, E. C. Spenser, J. Eckert, J. A. K. Howard, O. M. Yaghi. *Science*, *309*, 1350 (**2005**).

39. B. Chen, D. S. Contreras, N. W. Ockwig, O. M. Yaghi, *Angew. Chem. Int. Ed*, *44*, 4745 (2005).

40. H. Chun, D.N. Dybtsev, H. Kim, K. Kim, *Chem. Eur. J*, 11, 3521 (2005).

41. M. Dincă, A.F. Yu, J. R. Long, *J. Am. Chem. Soc*, 128, 8904 (2006).

42. S.S. Han, W.-Q. Deng, W.A. Goddard, *Angew. Chem. Int. Ed,* 46, 6289 (2007).

43. Z.X. Lian, J.W. Cai, C.H. Chen, *Polyhedron*, 26, 2647 (2007).

44. Z. Li, M. Li, X.P. Zhou, T. Wu, D. Li, S.W. Ng, *Cryst. Growth. Des*, 7, 1992 (2007).

45. M.P. Suh, Y.E. Cheon, E.Y. Lee, *Coord. Chem. Rev*, 252, 1007 (2008).

46. M. Dincă, A. Dailly, C. Tsay, J. R. Long, *Inorg. Chem*, 47, 11 (2008).

47. A. Mavrandonakis, E. Tylianakis, A.K. Stubos, G.E. Froudakis, *J. Phys. Chem. C,* 112, 7290 (2008).

48. K.L. Mulfort, J.T. Hupp, *Inorg. Chem,* 47, 7936 (2008).

49. S. Yang, X. Lin, A. J. Blake, K.M. Thomas, P. Hubberstey, N.R. Champness, M. Schroder, *Chem. Commun*, 46, 6108 (2008).

50. E. Klontzas, A. Mavrandonakis, E. Tylianakis, G.E. Froudakis, *Nano Lett*, 8, 1572 (2008).

51. A. Mavranodonakis, E. Klontzas, E. Tylianakis, G.E. Froudakis*, J. Am. Chem. Soc*, 131, 13410 (2009).

52. F. Nouar, J. Eckert, J.F. Eubank, P. Froster, M. Eddaoudi, *J. Am. Chem. Soc,* 131, 2864 (2009).

53. D. Himsi, D. Wallacher, M. Hartmann, *Angew. Chem. Int. Ed*, 48, 4639 (2009).

54. T.A. Maark, S. Pal, *Int. J. Hydrogen Energ*, 35, 12846 (2010).

55. X.L. Zou, M.H. Cha, S. Kim, M.C. Nguyen, G. Zhou, W.H. Duan, J. Ihm, Int. *J. Hydrogen Energ*, 35, 198 (2010).

56. N. S. Venkataramanan, R. Sahara, H. Mizuseki, Y. Kawazoe, *Inter. J. Mol. Sci*, 10, 1601 (2009).

57. R. Ströbel, J. Garche, P.T. Moseley, L. Jörissen, G. Wold, *J. Power Sources*, 159, 781 (2006).

58. J.C. Amicangelo, P.B. Armentrout, *J. Phys. Chem. A*, 104, 11420 (2000).

59. S.S. Han, H. Furukawa, O.M. Yaghi, W.A. goddard, *J. Am. Chem. Soc*, 130, 11580 (2008).

60. A. P. Cote, A. I. Benin, N.W. Ockwig, M. O'Keefee, A.J. Matzger, O.M. Yaghi, *Science*, 310, 1166 (2005).

61. H. M. El-Kaderi, J.R. Hunt, J.L. Mendoza-Cortes, A.P. Cote, R.E. Taylor, M. O'Keeffe, O.M. Yaghi, *Science*, 316, 268 (2007).

62. A. P. Cote, H.M. El-Kaderi, H. Furukawa, J.R. Hunt, O.M. Yaghi, *J. Am. Chem. Soc*, 129, 12914 (2007).

63. F.J. Uribe-Romo, J.R. Hunt, H. Furukawa, C. Klock, O'Keefee, O.M. Yaghi, *J. Am. Chem. Soc,* 131, 4570 (2009).

64. C.J. Doonan, D.J. Tranchemontagne, T.G. Glover, J.R. Hunt, O.M. Yaghi, *Nat. Chem*. 2, 235 (2010).

65. J.L. Mendoza-Cortes, S.S. Han, H. Furukawa, O.M. Yaghi, W. A. Goddard, *J. Phys. Chem. A*. 114, 10824 (2010).

66. N.N. McKeown, *J. Mater. Chem*. 20, 10588 (2010).

67. R.W. Rilford, S.J. Mugavero, P.J. Pellechia, J. Lavigne, *Adv. Mater*. 20, 2741 (2008).

68. M. Mastalerz, *Angew. Chem. Int. Ed*. 47, 445 (2008).

69. A. Thomas, *Angew. Chem. Int. Ed*. 49, 8328 (2010).

70. M.M. Wu, Q. Wang, Q. Sun, P. Jena, Y. Kawazoe, *J. Chem. Phys*, 133, 154706 (2010).

71. G. Garberoglio, *Langmiur*, 23, 12154 (2007).

72. N.S. Venkataramanan, M. Khazaei, R. Sahara, H. Mizuseki, Y. Kawazoe, *Chem. Phys,* 359, 173 (2009).

73. C. Janiak, J.K. Vieth, *New J. Chem*, 34, 2366 (2010).

74. E. Tylianakis, E. Klontzas, G.E. Froudakis, *Nanotechnology*, 20, 204030 (2009).

75. F. Svec, J. Germain, J.M.J. Frechet, *Small*, 5, 1098 (2009).

76. R. Babarao, J.W. Jiang, *Energ. Environ. Sci,* 1, 139 (2008).

77. Y.J. Choi, J.W. Lee, J.H. Choi, J.K. Kang, *Appl. Phys. Lett*, 92, 173102 (2008).

78. B. Assfour, G. Seifert, *Chem. Phys. Lett*, 489, 86 (2010).

79. D. Kim, D. H. Jung, S.-H. Choi, *J. Korean Phys. Soc,* 52, 1255 (2008).

80. Y.Y. Sun, K. Lee, Y.-H. Kim, S.B. Zhang, *Appl. Phys. Lett,* 95, 033109 (2009).

81. M. Suri, M. Dornfeld, E. Ganz, *J. Chem. Phys,* 131, 174703 (2009).

82. X. Zou, G. Zhou, W. Duan, K. Choi, J. Ihm, *J. Phys. Chem. C,* 114, 13402 (2010).

83. B. Engquist, P. Lötstedt, O. Runborg, Multiscale modeling and simulation in science, Springer-Verlag Berlin Heidelberg (2009).

84. J. Leszczynski, M. K. Shukla, Practical aspects of computational Chemistry, Springer Heidelberg Dordrecht (2009).

85. J. Lan, D. Cao, W. Wang, *Langmuir*, 26, 220 (2010).

86. J. Lan, D. Cao, W. Wang, B. Smit, *ACS Nano*, 4, 4225 (2010).

87. B.R. Babarao, J.W. Jiang, *Ind. Eng. Chem. Res*, 50, 62 (2010).

88. Y.H. Jin, B.A. Voss, R.D. Noble, W. Zhang, *Angew. Chem. Int. Ed,* 49, 6348 (2010).

89. G. Gareroglio, R. Vallauri, *Micropor. Mesopor. Mat*, 116, 540 (2008).

90. S.S. Han, H. Furukawa, O.M. Yaghi, W. A. Goddard, *J. Am. Chem. Soc,* 130, 11580 (2008).

91. E. Klontzas, E. Tylianakis, G.E. Froudakis, *J. Phys. Chem. C,* 112, 9095 (2008).

92. E. Klontzas, E. Tylianakis, G.E. Froudakis, *J. Phys. Chem. C,* 113, 21253 (2009).

93. E. Klontzas, E. Tylianakis, G.E. Froudakis, *Nano Lett,* 10, 452 (2010).

94. J. Lan, D. Cao, W. Wang, *J. Phys. Chem. C,* 114, 3108 (2010).

95. J. Lan, D. Cao, W. Wang, T. Ben, G. Zhu, *J. Phys. Chem. Lett,* 1, 978 (2010).

96. T. Tozawa, J.T.A. Jones, S.I. Swamy, S. Jiang, D.J. Adams, S. Shakespeare, R. Clowes, D. Bradshaw, T. Hasell, S. Y. Chong, C. Tang, S. Thompson, J. Parker, A. Trewin, J. Bacsa, A. M. Z. Slawin, A. Steiner, A.I. Cooper, *Nat. Mater*. 8, 973 (2008).

97. J.R. Holst, A. Trewin, A.I. Cooper, *Nat. Chem*. 2, 915 (2010).

98. P.K. Thallapally, B.P. McGrail, S.J. Dalgarno, H.T. Schaef, J. Tian, J.L. Atwood, *Nat. Mater*. *7*, 146 (2008)

99. P.K. Thallapally, G.O. Lloyd, T.B. Wirsig, M.W. Bredenkam, J.L. Atwood, L.J. Barbour, *Chem. Commun*. 5272 (2005).

100. J.L. Daschbasch, X. Sun, P.K. Thallapally, B.P. McGrail, L.X. Dang, *J. Phys. Chem. B,* 114, 5764 (2010).

101. J.L. Daschbach, P.K. Thallapally, J.L. Atwood, B.P. Mc Grail, L.X. Dang, *J. Chem. Phys*. *127*, 104703 (2007).

102. S. Alavi, J.A. Ripmeester, *Chem. Eur. J*. *14*, 1965 (2008).

103. N.S. Venkataramanan, R. Sahara, H. Mizuseki, Y. Kawazoe, *J. Phys. Chem. C*, 112, 19676 (2008).

104. N.S. Venkataramanan, R. Sahara, H. Mizuseki, Y. Kawazoe, *Comput. Mater. Sci,* 48, S263 (2010).

105. Blomqvist, A.; Araújo, C.M.; Srepusharawoot, P.; Ahuja, R. *PNAS*, **2007**, *104*, 20173.

106. A. Ranjbar, M. Khazaei, N.S. Venkataramanan, H. Lee, Y. Kawazoe, *Phys. Rev. B.* (In press).

107. M. Khazaei, M.S. Bahramy, N.S. Venkataramanan, H. Mizuseki, Y. Kawazoe, *J. Appl. Phys*. 106, 094303 (2009).

108. Y. Xue, G.A. Mansoori, *Int. J. Mol. Sci.* 11, 288 (2010).

**Figures Captions:**

1. A large series of isoreticular metal–organic frameworks (IRMOFs) has been produced in which each member shares the same cubic topology.

2. Primitive unit cell of isoreticular MOF-5 ( M = Fe, Co, Ni, Cu, Zn). Colors: blue (Metal center), gray (C atoms), white (H atoms), green (Li atom), red (oxygen). The structure is formed by OM$_4$ tetrahedra at the corners linked by benzene dicarboxylic (BDC) groups.

3. (a) Electrostatic potential spatial profile diagram for the Li-benzene system with H$_2$ physisorbed (b) charge density difference between the Li-benzene and Li-benzene with H$_2$ physisorbed on it.

4. Cluster unit used to understand the reactivity of Li-functionalized COF with dihydrogen molecule.

5. Optimized structure of TBC and four configuration of LiTBC. The relative energy $\Delta E$Li is evaluated by referring to configuration d.

6. Optimized geometries: (a) side view of calixarene with one hydrogen molecule, (b) on-top view of calixarene with one hydrogen molecule, and (c) side view of calixarene with two hydrogen molecules.

7. Pair distribution functions (PDF) from ab initio molecular dynamics simulations: (a) PDF for Li-benzene in the LTBC system, (b) PDF for Li-benzene in LTBC with four hydrogen molecules inside the cavity, and (c) PDF for Li-$H_2$ molecules in LTBC with four hydrogen molecules inside the cavity.

8. Excess (red)- depletion (blue) charge iso-surfaces and total and projected densities of states for (a) ADM.Li-$(H_2)_1$ and (b) ADM.Li-$(H_2)_4$.

9. Induced electric field in the direction of the Li/Li+−C bond of ADM.Li/Li+. Li/Li+ located at the origin. The dotted lines indicate the center of the adsorbed hydrogen molecule.

10. Occupation number as a function of the pressure and temperature on (a) ADM.Li and (b) ADM.Li+.

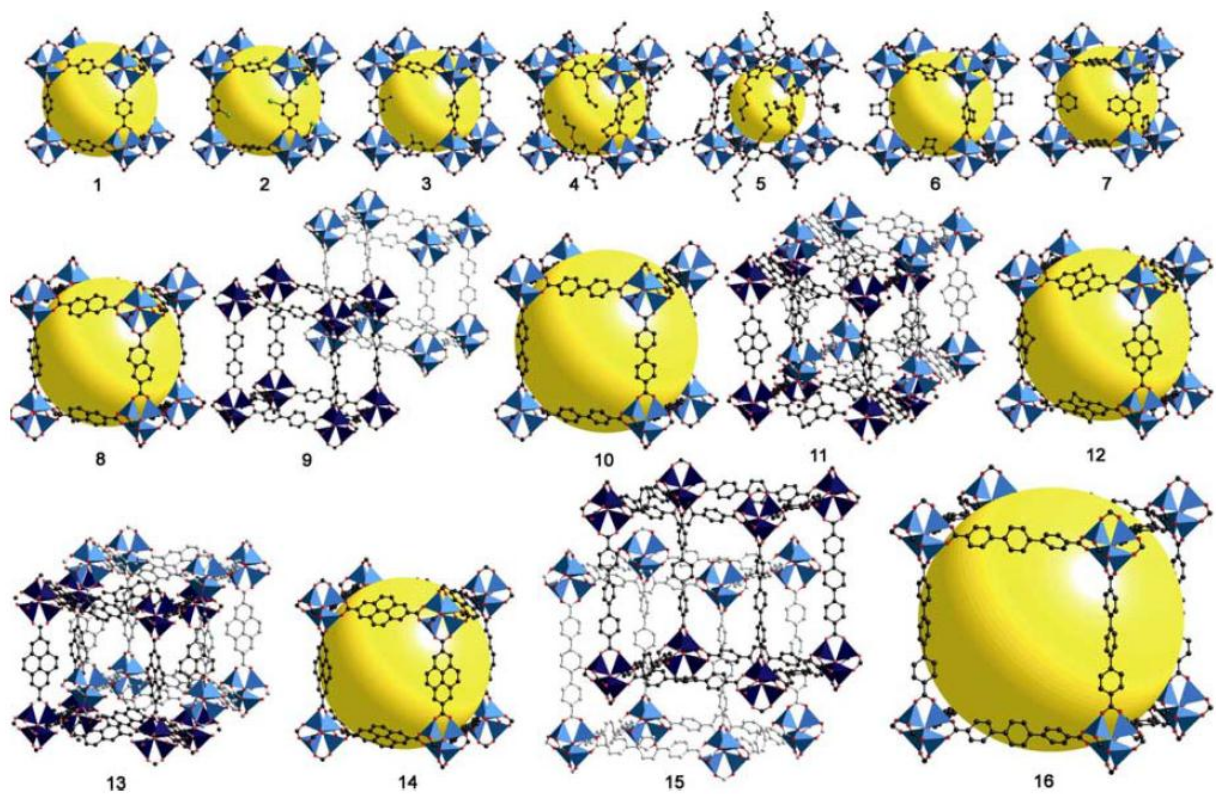

**Figure 1**

**Figure 2**

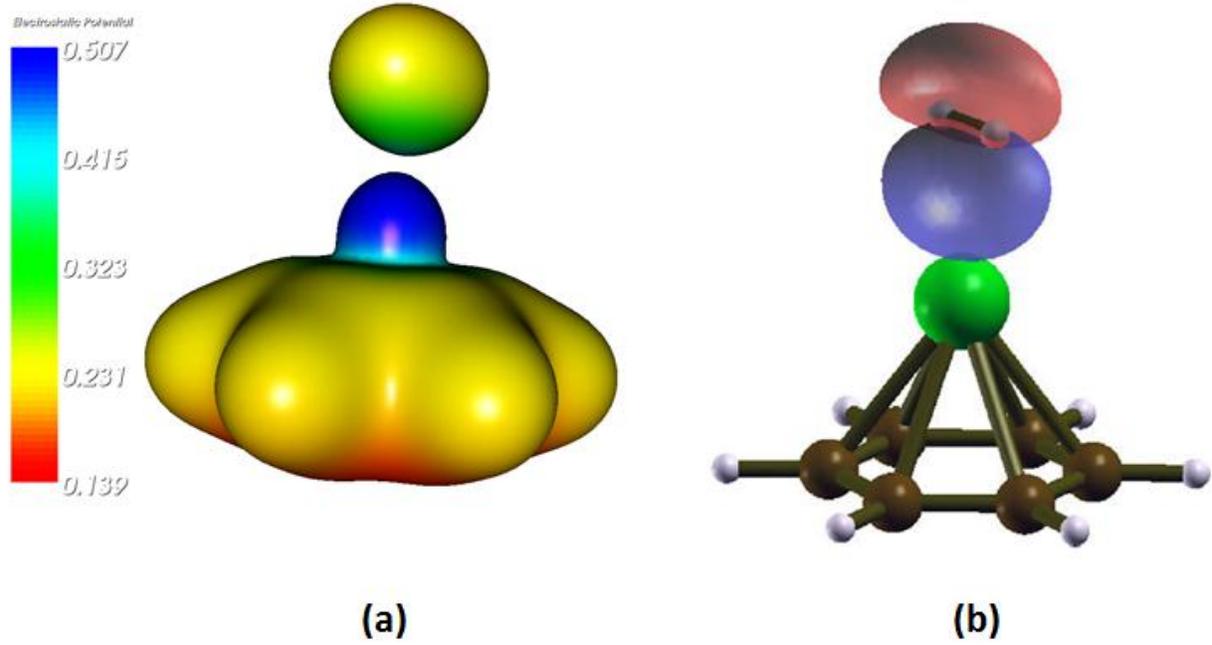

**Figure 3**

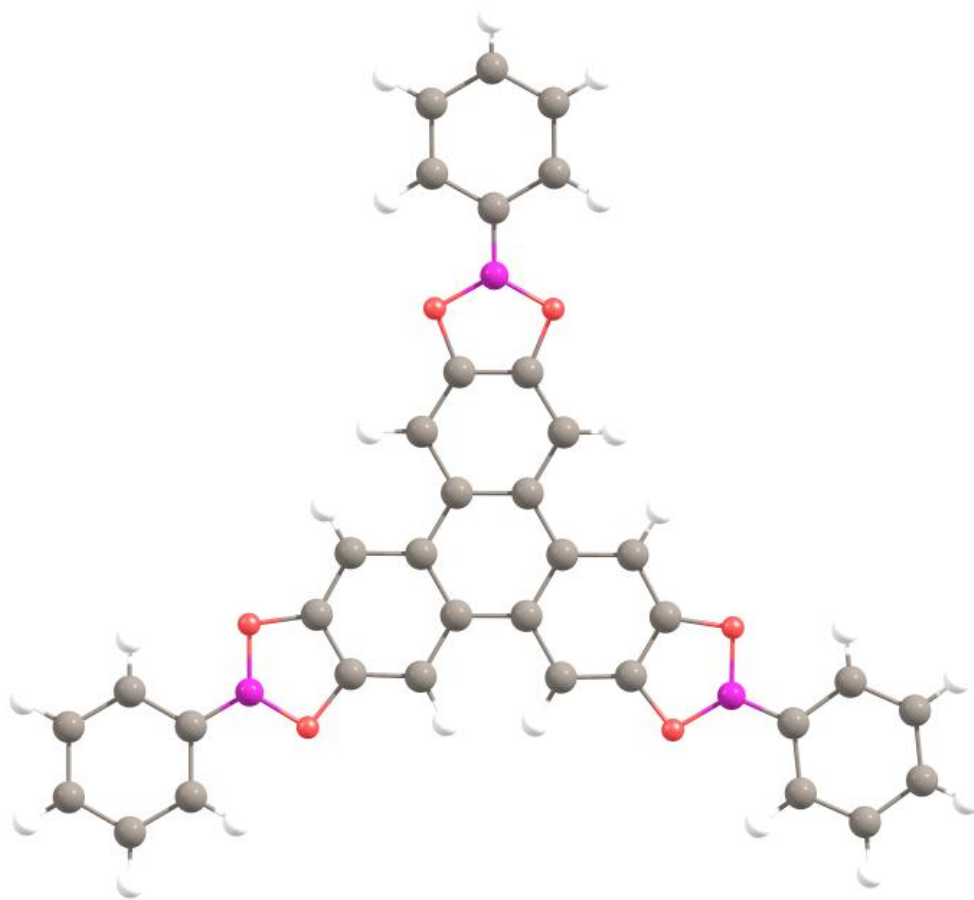

**Figure 4**

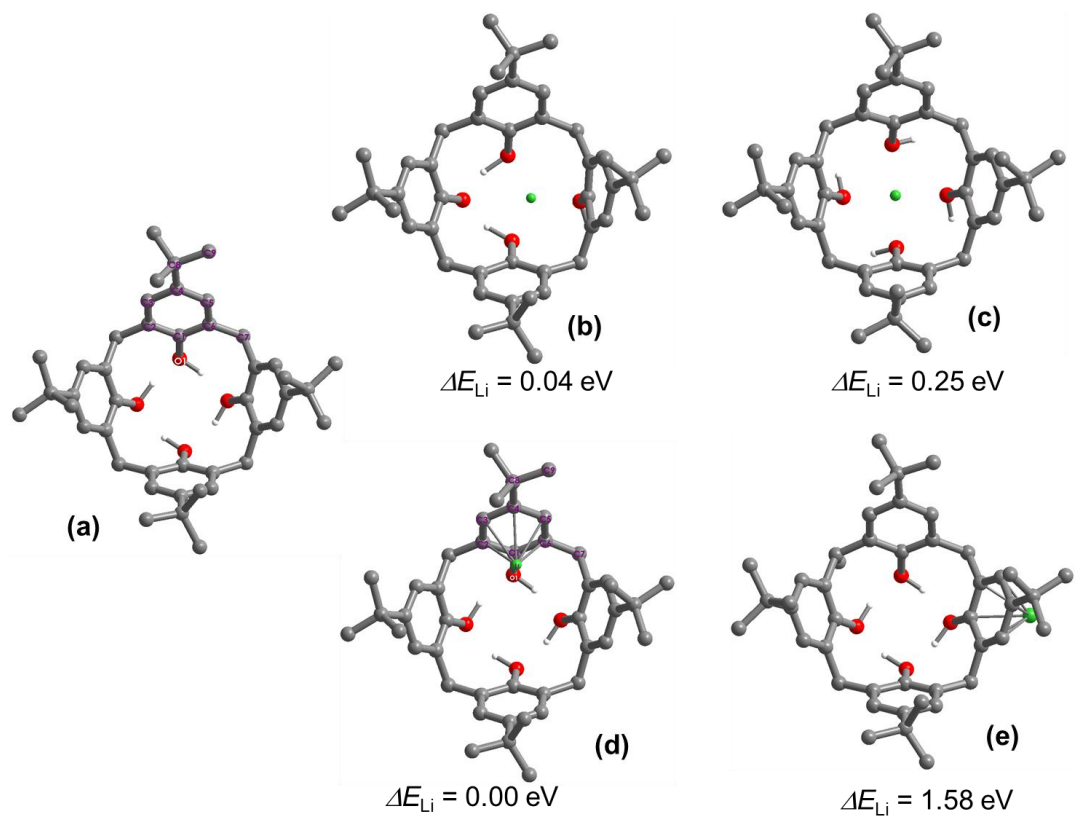

**Figure 5**

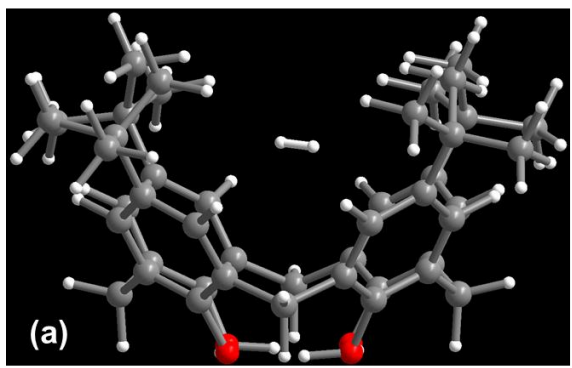
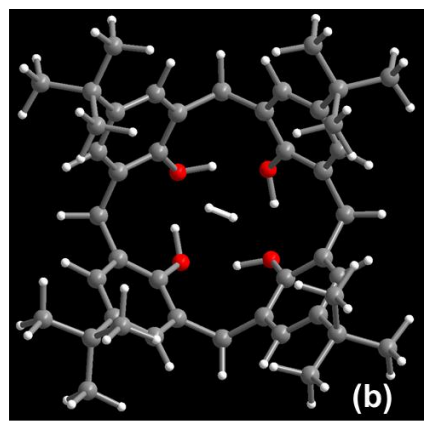
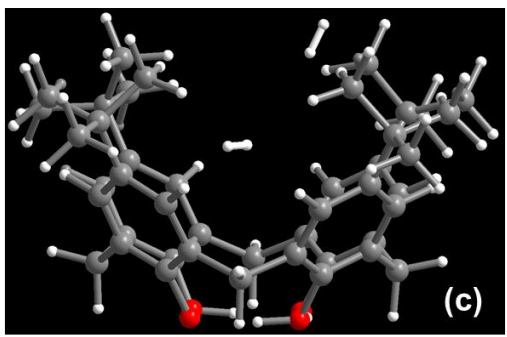

**Figure 6**

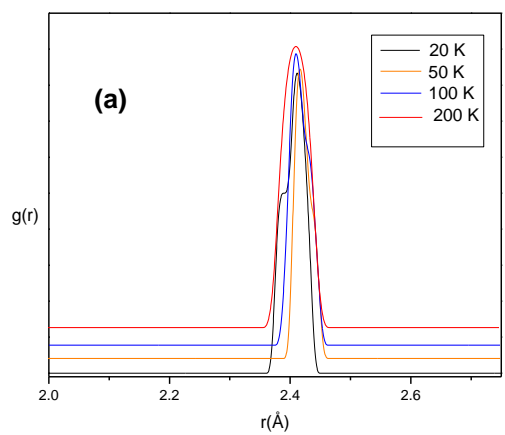
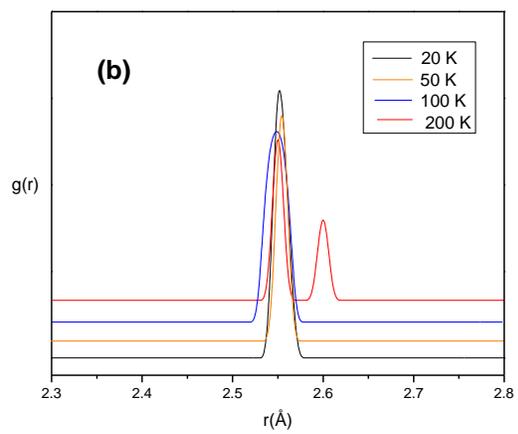
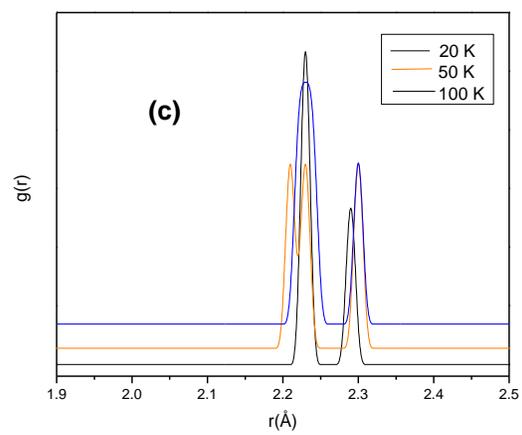

**Figure 7**

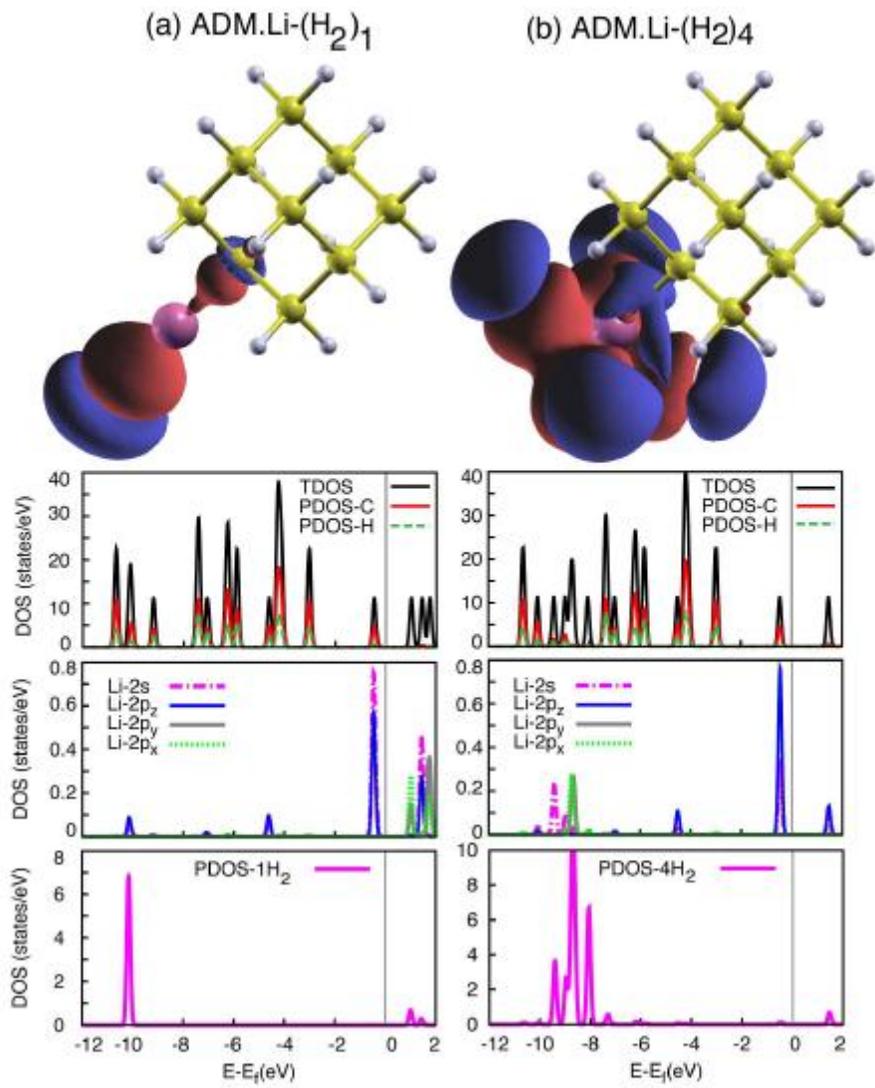

**Figure 8**

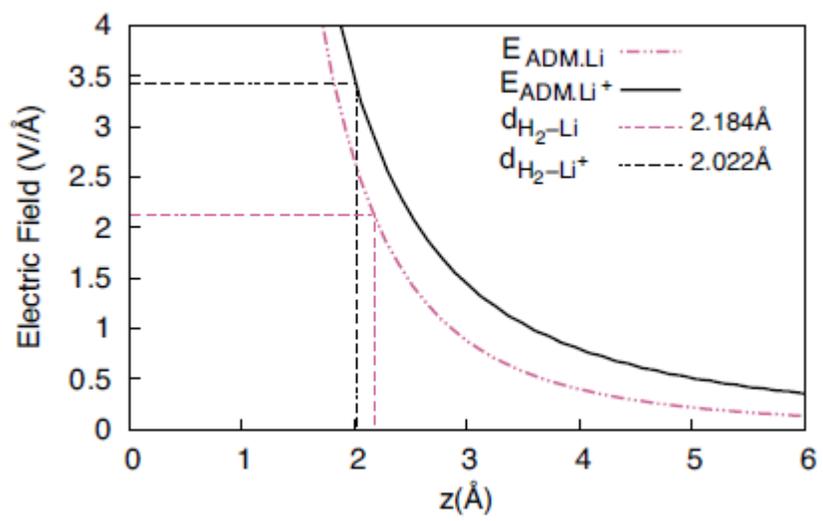

**Figure 9**

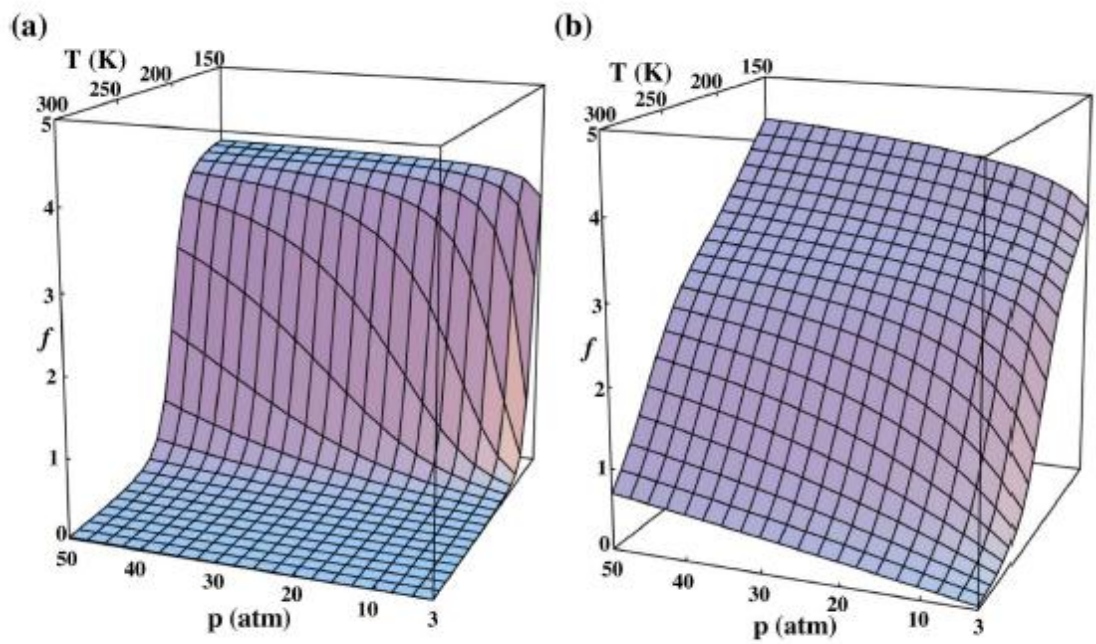

**Figure 10**

**Table 1:** Adsorption energy (AE, eV) and selected bond parameters (Å) for Li atom on Zn – MOF-5 unit.

| System | AE, eV | C2-C3[a] | C3-C4 | C4-C5 | C5-C6 | C6-C7 | C7-C2 | $C_M$-Li[b] |
|---|---|---|---|---|---|---|---|---|
| Li-Zn-MOF | -1.56 | 1.428 | 1.380 | 1.428 | 1.428 | 1.380 | 1.428 | 2.229 |
| Zn-MOF | | 1.399 | 1.384 | 1.399 | 1.399 | 1.384 | 1.399 | |

[a]For numbering see Figure 2.

[b]Mean Li-C distance

**Table 2**: Selected bond parameters (Å) and binding energy per hydrogen molecule (eV) for the adsorption of hydrogen on Li functionalized Zn-MOF.

| No. of $H_2$ | Avg. Benzene –Li (Å) | Li – $H_2$ (Å) | Avg. H – H (Å) | $\Delta E_b$ (eV) |
|---|---|---|---|---|
| 0 | 2.206 | – | 0.750 | - |
| 1 | 2.223 | 2.096 | 0.760 | 0.213 |
| 2 | 2.241 | 2.124 | 0.759 | 0.209 |
| 3 | 2.257 | 2.315 | 0.755 | 0.196 |
| 4 | 2.252 | 2.379 | 0.755 | 0.163 |

**Table 3.** Calculated structural parameters (Å) in the organic linker of Li cation doped M-MOF-5 (M = Fe, Co, Ni, Cu, Zn).

| System | C2-C3 | C3-C4 | C4-C5 | C5-C6 | C6-C7 | C7-C2 | $C_M$-Li[a] |
|---|---|---|---|---|---|---|---|
| Fe-MOF-5 | 1.425 | 1.369 | 1.425 | 1.425 | 1.396 | 1.425 | 2.274 |
| Co-MOF-5 | 1.417 | 1.407 | 1.417 | 1.417 | 1.407 | 1.417 | 2.292 |
| Ni-MOF-5 | 1.410 | 1.391 | 1.410 | 1.410 | 1.391 | 1.410 | 2.280 |
| Cu-MOF-5 | 1.407 | 1.396 | 1.408 | 1.407 | 1.394 | 1.404 | 2.285 |
| Zn-MOF-5 | 1.436 | 1.377 | 1.437 | 1.437 | 1.377 | 1.436 | 2.209 |

[a] Mean benzene – Li distance

**Table 4.** Selected bond lengths (Å) bond angles (deg) and adsorption energy of Li on MMOF-5 (M = Fe, Co, Ni, Cu, Zn).

| System | M-O1 | C1-O1 | C1-C2 | O1-C-O2 | M1-M2 | $\Delta AE$ eV |
|---|---|---|---|---|---|---|
| Fe-MOF-5 | 1.915 | 1.289 | 1.479 | 123.6 | 2.881 | 0.621 |
| Co-MOF-5 | 2.010 | 1.276 | 1.527 | 125.4 | 2.399 | 0.548 |
| Ni-MOF-5 | 1.949 | 1.270 | 1.481 | 127.1 | 2.864 | 0.776 |
| Cu-MOF-5 | 2.009 | 1.265 | 1.501 | 128.2 | 2.939 | 3.14 |
| Zn-MOF-5 | 1.956 | 1.264 | 1.491 | 129.5 | 3.128 | 1.15 |
| Zn-MOF-5[a] | 1.937 (1.911) | 1.274 (1.300) | 1.479 | 129.3 (125.0) | 3.128 (3.160) | - |

[a] Calculated and experimental values (in parenthesis) for Zn-MOF-5 without Li doping

**Table 5.**: Binding Energies (BE) for One to Four Hydrogen Molecules Absorbed on TBC and LTBC and average H–H Bond Distance

| No. of hydrogen molecule | TBC | | LTBC | |
|:---:|:---:|:---:|:---:|:---:|
| | BE/$H_2$ (eV) | H–H (Å) | BE/$H_2$ (eV) | H–H (Å) |
| 1 $H_2$ | 0.168 | 0.750 | 0.292 | 0.758 |
| 2 $H_2$ | 0.132 | 0.750 | 0.248 | 0.756 |
| 3 $H_2$ | – | – | 0.232 | 0.754 |
| 4 $H_2$ | – | – | 0.215 | 0.751 |

**Table 6**: $E_b$, the calculated consecutive binding energies (per Li) of Li /Li+ to ADM.Lim/Li$^+$ complexes. This binding energy is estimated as Eb=EADM.Lim+1-EADM.Lim-ELi. Single point MP2 calculations were done for the optimized structures obtained from M05-2X/6-311+G(2df,p).

| Clusters | PW91[a] | | PBE[a] | M05-2X[b] | | M05-2X[c] | MP2[d] |
|---|---|---|---|---|---|---|---|
| | $E_b$ | $d_{Li-C}$ | $E_b$[a] | $E_b$ | $d_{Li-C}$ | $E_b$ | $E_b$ |
| ADM.Li | -1.46 | 2.028 | -1.29 | -1.57 | 2.010 | -1.51 | -1.59 |
| ADM.Li2 | -1.16 | -2.050 | -1.02 | -1.36 | 2.023 | -1.30 | -1.36 |
| ADM.Li3 | -0.94 | 2.075 | -0.84 | -1.17 | 2.041 | -1.14 | -1.24 |
| ADM.Li4 | -0.71 | 2.094 | -0.62 | -1.06 | 2.059 | -1.05 | -1.25 |
| ADM.Li+ | -1.62 | 2.082 | -1.60 | -1.52 | 2.038 | -1.46 | -1.43 |

[a]Basis set: plan-wave.
[b]DFT cluster method, basis set: 6-311+G(2df, p).
[c]DFT cluster method, basis set: 6-31+G(d, p).
[b]Basis set: 6-311+G(2df, p).

**Table 7.** $E_b$ (in eV), $d_{H-H}$, $d_{H2-Li}$, and $d_{Li-C}$ (in Å) are the calculated binding energies of adsorbed hydrogen molecules, bond length average of hydrogen molecule(s), average distance between the center of hydrogen molecule(s) and Li/Li$^+$, and average bond distance of Li and host carbon atoms in different methods, respectively.

| cluster | PW91 | | | | PBE | | M05-2X | | | M05-2X | MP2 |
|---|---|---|---|---|---|---|---|---|---|---|---|
| | $E_b$ | $d_{H-H}$ | $d_{H2-Li}$ | $d_{Li-C}$ | $E_b$ | $E_b$ | $d_{H-H}$ | $d_{H2-Li}$ | $d_{Li-C}$ | $E_b$ | $E_b$ |
| ADM.Li-(H$_2$)$_1$ | -0.10 | 0.753 | 2.184 | 2.020 | -0.10 | -0.11 | 0.743 | 2.149 | 2.008 | -0.11 | -0.11 |
| ADM.Li-(H$_2$)$_2$ | -0.20 | 0.788 | 1.788 | 2.007 | -0.20 | -0.15 | 0.760 | 1.868 | 1.948 | -0.12 | -0.10 |
| ADM.Li-(H$_2$)$_3$ | -0.15 | 0.770 | 1.882 | 2.049 | -0.14 | -0.15 | 0.749 | 1.965 | 2.011 | -0.14 | -0.12 |
| ADM.Li-(H$_2$)$_4$ | -0.14 | 0.763 | 1.966 | 2.088 | -0.13 | -0.15 | 0.748 | 2.034 | 2.035 | -0.14 | -0.12 |
| ADM.Li-(H$_2$)$_5$ | -0.11 | 0.760 | 2.285 | 2.085 | -0.11 | -0.13 | 0.746 | 2.197 | 2.023 | -0.13 | -0.10 |
| ADM.Li4-(H$_2$)$_{20}$ | -0.12 | 0.765 | 2.333 | 2.135 | -0.11 | -0.12 | 0.750 | 2.221 | 2.048 | -0.11 | -0.08 |
| ADM.Li$^+$-(H$_2$)$_1$ | -0.23 | 0.757 | 2.022 | 2.090 | -0.21 | -0.21 | 0.746 | 2.042 | 2.046 | -0.17 | -0.21 |
| ADM.Li$^+$-(H$_2$)$_2$ | -0.21 | 0.757 | 2.025 | 2.146 | -0.19 | -0.20 | 0.746 | 2.028 | 2.113 | -0.16 | -0.20 |
| ADM.Li$^+$-(H$_2$)$_3$ | -0.19 | 0.756 | 2.049 | 2.184 | -0.17 | -0.19 | 0.746 | 2.031 | 2.145 | -0.15 | -0.19 |
| ADM.Li$^+$-(H$_2$)$_4$ | -0.17 | 0.755 | 2.139 | 2.223 | -0.15 | -0.17 | 0.744 | 2.124 | 2.187 | -0.14 | -0.17 |
| ADM.Li$^+$-(H$_2$)$_5$ | -0.15 | 0.755 | 2.236 | 2.270 | -0.13 | -0.16 | 0.744 | 2.192 | 2.241 | -0.13 | -0.15 |
| H$_2$ | | 0.749 | | | | | 0.739 | | | | |

[a]Basis set: plan-wave ; [b]DFT cluster method, basis set: 6-311+G(2df, p) ; [c]DFT cluster method, basis set: 6-31+G(d, p) ; [d]Basis set: 6-311+G(2df, p).